\documentclass[superscriptaddress,PRD,twocolumn,reprint,nofootinbib]{revtex4-2}

\usepackage[utf8]{inputenc}
\usepackage{lmodern}
\usepackage{exscale}

\usepackage{amsmath}
\usepackage{amssymb}
\usepackage{mathtools}

\usepackage{mathrsfs}

\usepackage{relsize,exscale}
\usepackage{tensor}

\usepackage[low-sup]{subdepth}

\makeatletter
\newcommand{\dalembertian}{\mathop{\mathpalette\dalembertian@\relax}}
\newcommand{\dalembertian@}[2]{%
  \begingroup
  \sbox\z@{$\m@th#1\square$}%
  \dimen0=\fontdimen8
    \ifx#1\displaystyle\textfont\else
    \ifx#1\textstyle\textfont\else
    \ifx#1\scriptstyle\scriptfont\else
    \scriptscriptfont\fi\fi\fi3
  \makebox[\wd\z@]{%
    \hbox to \ht\z@{%
      \vrule width \dimen0
      \kern-\dimen0
      \vbox to \ht\z@{
        \hrule height \dimen0 width \ht\z@
        \vss
        \hrule height 2\dimen0
      }%
      \kern-2.5\dimen0
      \vrule width 2.5\dimen0
    }%
  }%
  \endgroup
}
\makeatother

\usepackage[normalem]{ulem}
\usepackage{cancel}

\usepackage{hyperref}
\hypersetup{
    colorlinks,
    linkcolor={black},
    citecolor={black},
    urlcolor={black}
}

\begin{document}

\title{Critical reassessment of the restricted Weyl symmetry}

\author{Dra\v{z}en Glavan}
\email[\tt e-mail: ]{glavan@fzu.cz}
\affiliation{CEICO, FZU --- Institute of Physics of the Czech Academy of Sciences, 
Na Slovance 1999/2, 182 21 Prague 8, Czech Republic}

\author{Ruggero Noris}
\email[\tt e-mail: ]{noris@fzu.cz}
\affiliation{CEICO, FZU --- Institute of Physics of the Czech Academy of Sciences, 
Na Slovance 1999/2, 182 21 Prague 8, Czech Republic}

\author{Tom Zlosnik,}
\email[\tt e-mail: ]{thomas.zlosnik@ug.edu.pl}
\affiliation{Institute of Theoretical Physics and Astrophysics, 
University of Gda\'{n}sk, 80-308 Gda\'{n}sk, Poland}

\begin{abstract}
A class of globally scale-invariant scalar-tensor theories have been
proposed to be invariant under a larger class of transformations that take
the form of local Weyl transformations supplemented by a 
restriction that the conformal factor satisfies a covariant 
Klein-Gordon equation. The action of these theories indeed seems
to be invariant under such transformations up to boundary terms,
this property being referred to as ``restricted Weyl symmetry''.
However, we find that corresponding equations of motion are not
invariant under these transformations. This is a paradox, that is explained by
realizing that the restriction condition on the conformal factor
forces the restricted Weyl transformation to 
be a {\it nonlocal} transformation. For nonlocal 
transformations would-be boundary terms cannot in general be discarded from the action.
Moreover, variations of trajectories cannot be assumed
to vanish at boundaries of the action when deriving equations of motion. 
We illustrate both of these less known properties by considering a series of simple examples.
Finally, we apply these observations to the case of globally scale-invariant scalar-tensor theories to demonstrate that restricted Weyl transformations are, in fact, not symmetries of the full system.
\end{abstract}

\maketitle

\section{Introduction and motivation}
\label{sec:IntroductionAndMotivation}

The dynamics of physical systems can exhibit invariance under two kinds
of transformations, referred to as global 
symmetries and local/gauge symmetries.
The former are characterized by constant (global) parameters, 
and map solutions of equations of motion
onto physically distinct solutions of the same equations. 
The latter, on the other hand, are a consequence of redundancy in the description
of dynamics, where the transformations, parametrized by an
arbitrary spacetime dependent function, map between solutions 
describing the same physical realization of the system. In terms of the initial value problem, global transformations map one
set of initial conditions into another set, thus relating the
evolution of the two systems, while local transformations
affect just the evolution without changing initial conditions,
thus describing one and the same system.
Still, proposals have been put forward for a concept of
another concept of symmetry that lies somewhere in between.

``Restricted symmetries''  are obtained by promoting a constant parameter of 
a global symmetry transformation to a spacetime dependent function.
However, rather than gauging this transformation, the local function is restricted to satisfy its own equation of motion. This restriction 
equation does not follow from the action principle, but rather from 
the requirement that the action remains invariant. 
In particular, this concept has been introduced
in~\cite{Edery:2014nha,Edery:2015wha} for globally scale-symmetric
scalar-tensor theories, inspired by earlier works on harmonic Weyl transformations~\cite{ORaifeartaigh:1996hvx,Iorio:1996ad}.
The global scale symmetry transformation is promoted to a local 
Weyl transformation with an additional restriction.
The conformal factor is required to satisfy a particular equation
that ensures the scalar invariants comprising the action transform 
homogeneously up to boundary terms, making the action invariant under the 
transformation.
For this reason these theories are inferred to be symmetric under such 
restricted Weyl transformations.
This idea has been investigated further and expanded upon in subsequent 
works~\cite{Edery:2018jyp,Edery:2023hxl,Oda:2020wdd,Kamimura:2021wzf,Oda:2021kvx,Oda:2020cmi,Edery:2021bbe}.
Most recently it has been revisited in~\cite{Martini:2024tie}, where
also other ``restricted subgroups'' of Weyl transformations,
that could serve as a basis for ``restricted symmetries'',
have been identified, including the Liouville transformation that
has been previously considered in~\cite{Kuhnel:1995}.
The concept of ``restricted symmetries'' could indeed have profound consequences,
and this is why it warrants closer inspection.

In order to make the discussion concrete, let us consider 
the following scalar-tensor theory given by the action
\begin{equation}
\!\! S [g_{\mu\nu},\phi ] =\! \int\! d^{4\!}x \, \sqrt{-g} \,
    \biggl[ \alpha R^2 
    - \frac{1}{2} \nabla^\mu\phi \nabla\! _\mu \phi
    - \frac{\xi}{2} R \phi^2 
    \biggr]
    .
\label{STaction}
\end{equation}
This action is symmetric under global scale transformations, 
i.e.~it is invariant under
simultaneous rescaling of the metric~$g_{\mu\nu}(x) \!\to\! \Omega^2 g_{\mu\nu}(x)$
and of the scalar field~$\phi(x) \!\to\! \phi(x)/\Omega$ by a 
constant parameter~$\Omega$.
Consequently, the associated equations of motion are also
left unchanged by this global transformation, both the scalar equation,
\begin{equation}
\dalembertian\phi-\xi R\phi=0 \, ,
\label{scalarEOM}
\end{equation}
and the tensor one,
\begin{align}
\MoveEqLeft[1]
4 \alpha \Bigl[ 
    \nabla_\mu \nabla_\nu 
    \!-\! 
    g_{\mu\nu} {\dalembertian}
    \!-\! 
    R_{\mu\nu} 
    \!+\! 
    \tfrac{1}{4} R g_{\mu\nu}\Bigr] R
\label{tensorEOM}
\\
&
    +  
    {{\mathbb P}_{\mu\nu}}^{\rho\sigma} 
        (\partial_\rho \phi ) (\partial_\sigma \phi)
    + 
    \xi
    \Bigl[ 
        G_{\mu\nu} 
        \!-\! 
        \nabla_\mu \nabla_\nu 
        \!+\! 
        g_{\mu\nu} {\dalembertian} \, \Bigr] \phi^2
        = 0
    \, ,
\nonumber 
\end{align}
where~$ \dalembertian \!=\! g^{\mu\nu} \nabla_\mu \nabla_\nu$ is the 
d'Alembertian operator, and~${{\mathbb P}_{\mu\nu}}^{\rho\sigma} 
    \!=\! \delta_{(\mu}^\rho \delta_{\nu)}^\sigma \!-\! \frac{1}{2} g_{\mu\nu} g^{\rho\sigma}$
is the trace-reversal projector.
The key step in~\cite{Edery:2014nha,Edery:2015wha} is to promote the
global scale transformation to a local one,
\begin{align}
g_{\mu\nu}(x) \longrightarrow{}&
    g_{\mu\nu}'(x) 
    = 
    \Omega^{-2}(x) g_{\mu\nu}(x) \, ,
\label{restrictedWeylTransMetric}
\\
\phi(x) \longrightarrow{}& 
    \phi'(x)
    =
    \Omega(x) \phi(x) \, ,
\label{restrictedWeylTrans}
\end{align}
with $\Omega(x)$ satisfying a covariant Klein-Gordon equation,
\begin{equation}
{\dalembertian}' \Omega (x) = 0 \, ,
\label{WeylRestriction}
\end{equation}
where a prime on operators denotes that they depend 
on the primed metric.
This transformation indeed leaves the action in~(\ref{STaction}) invariant up 
to discarding ``boundary terms'', and is thus considered 
a ``restricted Weyl symmetry''. 
The Klein-Gordon equation~(\ref{WeylRestriction}) exemplifies how ``restricted symmetries'' should differ from global and local symmetries. 
Its solutions are fully specified once initial conditions for~$\Omega$
are given. The freedom in their choice represents a functional freedom 
of the restricted local transformation, akin to freedom of local symmetries. 
However, nontrivial solutions for the conformal
factor~$\Omega$, that do not correspond to global transformations,
will necessarily fashion a Weyl transformation 
that alters initial conditions of the scalar-tensor theory, which is 
a property shared with global transformations.
In this sense the concept of 
``restricted symmetries'' is somewhere in between global and local 
symmetries.

One expects that if the action is symmetric under the transformation, 
then the equations of motion descending from the action principle
would share the same property. However, 
an unexpected feature is revealed when this 
is checked. While the scalar 
equation~(\ref{scalarEOM}) indeed is invariant under the
transformation, the tensor equation~(\ref{tensorEOM}) is not,
\begin{align}
\MoveEqLeft[1]
4 \alpha \Bigl[ 
    \nabla'_\mu \nabla'_\nu 
    \!-\! 
    g'_{\mu\nu} {\dalembertian}'
    \!-\! 
    R'_{\mu\nu} 
    \!+\! 
    \tfrac{1}{4} R' g'_{\mu\nu}\Bigr] R'
\nonumber \\
&  
    +
    {{\mathbb P}'_{\mu\nu}}^{\rho\sigma} 
        (\partial_\rho \phi' ) (\partial_\sigma \phi')
    +
    \xi
    \Bigl[ 
        G'_{\mu\nu} 
        \!-\! 
        \nabla'_\mu \nabla'_\nu 
        \!+\! 
        g'_{\mu\nu} {\dalembertian}' \, \Bigr] \phi'^2
\nonumber \\
&   \hspace{0.4cm}
    =
    {{\mathbb P}'_{\mu\nu}}^{\rho\sigma}
    \frac{ \partial_\sigma \Omega }{\Omega} 
    \Bigl[
    \partial_\rho
    \!-\!
    \frac{ \partial_\rho\Omega }{\Omega} 
    \Bigr]
    \Bigl[
    24\alpha R'
    \!+\!
    (1\!-\!6\xi) \phi'^2
    \Bigr]
    \, ,
\label{IntroNoninvariance}
\end{align}
even with the restriction condition accounted~(\ref{WeylRestriction}) for.
This paradoxical result, that seems to have gone unnoticed
thus far, makes it difficult to argue in favour of 
``restricted Weyl symmetry'' as being a symmetry. 
The resolution of the 
paradox lies in the conclusions pertaining to the properties of the action, 
under the transformation~(\ref{restrictedWeylTrans}) with the 
restriction~(\ref{WeylRestriction}).

This work is therefore devoted to the critical assessment of the concept of ``restricted symmetries''. To this end, in Secs.~\ref{sec: First example}--\ref{sec: Fourth example}, 
we examine a series of simple particle mechanics examples, which nonetheless capture the 
key properties giving rise to the issues observed here.
We show that the transformations involved in ``restricted symmetries''
are generally {\it nonlocal} transformations. This follows from restriction
conditions such as~(\ref{WeylRestriction}), which force 
the transformation function to be a nonlocal function of
dynamical variables. 
In all the examples considered, we find that
this feature causes the noninvariance of the equations of motion. 
Accounting for the nonlocal nature of the transformation at the level of the action
required a better understanding of boundary terms and boundary variations,
neither of which can be neglected in general.
In Sec.~\ref{sec: Scalar-tensor theory} we apply these lessons to nonlinear 
perturbations of the scalar-tensor theory~(\ref{STaction}) 
around Minkowski space, and resolve the paradox observed in this introduction,
concluding that the ``restricted Weyl symmetry'' is neither a local
transformation, nor a symmetry transformation.

\section{First example: linear superposition}
\label{sec: First example}

We start the series of examples by considering arguably the simplest system possible, that of a free particle on a line. Its dynamics is encoded by the action
\begin{equation}
S[\phi] = \int_{t_a}^{t_b}\! dt \, \frac{\dot{\phi}^2}{2} \, ,
\label{ONEaction}
\end{equation}
where $t_a$ and $t_b$ are initial and final time, respectively, and where henceforth an overdot denotes 
a time derivative. 
This action is symmetric under global shifts of the position variable,
$\phi(t) \!\to\! \phi(t) \!+\! s $, and consequently
so is the associated equation of motion,
\begin{equation}
\ddot{\phi} = 0 \, .
\end{equation}

Let us now consider promoting the global scale transformation to a 
local one,
\begin{equation}
\phi(t) \longrightarrow \varphi(t) = \phi(t) - s(t) \, .
\label{ONCElocalTrans}
\end{equation}
The action~(\ref{ONEaction}) will not be invariant for
an arbitrary transformation function~$s(t)$, but rather transforms
to
\begin{equation}
S[\phi] \quad\longrightarrow\quad S[\varphi]
    =
    \int_{t_a}^{t_b}\! dt \, 
    \biggl[
    \frac{\dot{\varphi}^2}{2}
    +
    \dot{s}
    \Bigl( \dot{\varphi} + \frac{\dot{s}}{2} \Bigr)
    \biggr]
    \, .
\label{ONEactionTrans}
\end{equation}
Only if there is a condition that allows us to discard the additional
term generated by the transformation would the action be invariant.
There are several options; taking~$\dot{s} \!=\! 0$ ensures 
that the extra term vanishes, but just brings us back to the global 
shift symmetry;
taking~$\dot{s}\!=\! - 2 \dot{\varphi}$ also eliminates the extra 
term but this is just a convoluted manifestation of the 
discrete~$\mathbb{Z}_2$ symmetry~$\phi \!\to\! - \phi$ of the 
original action~(\ref{ONEaction}).

If we allow for partial integration 
and throwing away boundary terms, then restricting the transformation function~$s(t)$ 
to satisfy
\begin{equation}
\ddot{s} = 0 \, 
\label{ONErestriction}
\end{equation}
will guarantee that the action~(\ref{ONEaction}) is invariant 
under such a restricted local shift transformation. The equation of motion descending from the transformed action~(\ref{ONEactionTrans}) matches the one obtained by 
transforming the original equation~(\ref{ONEaction}),
\begin{equation}
\ddot{\varphi} + \ddot{s} = 0 \, .
\end{equation}
It is clear that this equation is invariant under the local  
shift transformation~(\ref{ONCElocalTrans}),
provided that the restriction in~(\ref{ONErestriction})
holds. 
However, we should not conclude the transformation 
in~(\ref{ONEactionTrans}) to be
a new type of restricted symmetry.
In fact, what we had uncovered here is just the superposition principle, valid for any homogeneous system of linear equations:
a linear combination of two solutions of such equations is 
again a solution. To properly examine the concept of restricted 
transformations, we need to consider nonlinear systems, which we do
in the rest of the examples.

\section{Second example: shift symmetry}
\label{sec: Second example}

Having concluded in the preceding section that restricted symmetries in linear
systems are nothing but the linear superposition principle, here we move on
to consider nonlinear theories. In particular, we will analyse a 
simple nonlinear two-particle mechanical system exhibiting
global shift symmetry, given by the action

\begin{align}
S[\rho,\phi] = \int_{t_a}^{t_b}\! dt \, 
    \biggl[
    \frac{\dot{\rho}^2}{2}
    + \frac{\rho^2 \dot{\phi}^2 }{2}
    - U(\rho)
    \biggr]
    \, .
\label{1Daction}
\end{align}
The equations of motion descending from the action principle
applied to~(\ref{1Daction}) are
\begin{align}
    &
    \ddot\rho-\rho\dot\phi^2 + U'(\rho) = 0
    \, ,
\label{EOMrho}
\\
    &
    \frac{d}{dt} \bigl( \rho^2\dot\phi \bigr) = 0
    \, .
\label{EOMphi}
\end{align}
Both the action in~(\ref{1Daction}) and, consequently, the equations of motion above
are invariant under global shifts of one of the 
fields,~$\phi(t) \!\to\! \phi(t) \!+\! f$. Let us now try to promote this global symmetry to a restricted local symmetry
\begin{equation}
\phi(t) \longrightarrow \varphi(t) = \phi(t) - f(t) \, ,
\label{localTrans}
\end{equation}
where~$\varphi(t)$ is the new dynamical field, and~$f(t)$ is 
the transformation function
satisfying some restriction condition yet to be determined. 
In general, the action in~(\ref{1Daction}) is not invariant under this
transformation for an arbitrary
function~$f(t)$, but rather it transforms to
\begin{align}
S[\rho,\phi] \quad \longrightarrow \quad
S[\rho,\varphi] ={}&
    \int_{t_a}^{t_b} \! dt \, \biggl[
    \frac{\dot{\rho}^2}{2}
    + \frac{\rho^2 \dot{\varphi}^2 }{2}
    - U(\rho)
\nonumber \\
&
    +
    \frac{\rho^2 \dot{f}}{2} \bigl( 2 \dot{\varphi} + \dot{f} \bigr)
    \biggr]
    \, .
\label{DS}
\end{align}
Requiring that the additional term in
the second line above vanishes identically would result in the 
action not transforming under~(\ref{localTrans}). However, just as in Sec.~\ref{sec: First example},
this requirement would only produce trivial restrictions of~$f(t)$:
(i) we may either require~$\dot{f} \!=\! 0$, which only recovers global shift symmetry,
or (ii) we may require~$\dot{f} \!=\! - 2 \dot{\varphi}$, which is just a complicated way of writing a discrete~$\mathbb Z_2$ transformation~$\phi(t) \!\to\! - \phi(t)$.

Obtaining a nontrivial restriction on~$f(t)$, if it exists, necessarily requires further manipulations, and the only one we have at our disposal is partial
integration. Led by the usual expectation that boundary terms of the action do not contribute to bulk equations of motion, we partially integrate~(\ref{DS}) into the form
\begin{equation}
\Delta S[\rho, \varphi] = \int_{t_a}^{t_b} \! dt \, 
    \biggl[
    - \frac{1}{2} \bigl( 2\varphi + f \bigr) \frac{d}{dt}\bigl( \rho^2 \dot{f} \bigr)
    \biggr]
    +
    {\tt (b.t.)}
    \, ,
\label{secondDS}
\end{equation}
where~$\tt (b.t.)$ stands for boundary terms.
Now, requiring the action to be invariant {\it up to boundary terms} 
produces a restriction on the transformation~(\ref{localTrans}),
\begin{equation}
\frac{d}{dt} \bigl( \rho^2 \dot{f} \bigr) = 0 \, ,
\label{fRestriction}
\end{equation}
which does not reduce to global symmetry transformations.
It should be noted that this restriction equation should be satisfied
off-shell, meaning for an arbitrary field~$\rho(t)$. Given that
only boundary terms in~(\ref{secondDS}) remain, the transformed action
generates the same equations of motion as~(\ref{EOMrho}) 
and~(\ref{EOMphi}) with~$\varphi$ in place of~$\phi$.

The common lore would now have it that if the action remains invariant 
under a transformation (up to boundary terms), 
so do the equations of motion. However,
a simple check reveals this not to be the case. Applying the transformation (\ref{localTrans}), together with the restriction (\ref{fRestriction}), directly to the original equations of motion~(\ref{EOMrho}) and~(\ref{EOMphi}) reveals them to, respectively, transform to
\begin{align}
& 
    \ddot{\rho} - \rho \dot{\varphi}^2 + U'(\rho)
        = 2 \rho \dot{\varphi} \dot{f} + \rho \dot{f}^2
        \, ,
\label{rehoEqTrans1}
\\
& 
    \frac{d}{dt} \bigl( \rho^2 \dot{\varphi} \bigr) = 0
    \, .
\label{phiEqTrans1}
\end{align}
While the second equation does remains invariant, the first one 
receives additional contributions that depend on the transformation function~$f(t)$. 
Such behaviour, reminiscent of the one observed for the scalar-tensor theory
in Sec.~\ref{sec:IntroductionAndMotivation}, does not usually manifest for local 
transformations of dynamical variables. This warrants a closer examination
of the restriction condition~(\ref{fRestriction}), which is what we
turn our attention to next.

The differential equation~(\ref{fRestriction}), representing the restriction
condition on the transformation 
function~$f(t)$, is a local equation. However, the fact 
that it has to be satisfied off-shell for an arbitrary~$\rho(t)$
makes~$f(t)$ a nonlocal function (i.e.~a functional) of the dynamical 
variable~$\rho(t)$. We have chosen this example to be simple enough so that this functional dependence can be exhibited explicitly.
Equation~(\ref{fRestriction}) can be integrated twice to yield
\begin{equation}
f(t) = f_0 + c_0 \int_{t_0}^{t} \! \frac{dt'}{\rho^2(t')} \, ,
\label{fSol}
\end{equation}
where~$f_0$ and~$c_0$ are constants of integration (the former being just a manifestation of global shift symmetry), 
and~$t_0$ is some time,~$t_a \!<\! t_0 \!<\! t_b$. 
However, even after taking this solution into account, the right-hand side of~(\ref{rehoEqTrans1}) does not vanish. Rather, original equations of motion~(\ref{EOMrho}) and~(\ref{EOMphi}) are seen to, respectively, transform to
\begin{align}
& 
    \ddot{\rho} - \rho \dot{\varphi}^2 + U'(\rho)
        =
            \frac{2 c_0 \dot{\varphi} }{\rho}
        + 
        \frac{c_0^2}{\rho^3}
        \, ,
\label{rehoEqTrans2}
\\
& 
    \frac{d}{dt} \bigl( \rho^2 \dot{\varphi} \bigr) = 0
    \, ,
\label{phiEqTrans2}
\end{align}
and thus the transformation does not leave the equations of motion invariant.

The solution~(\ref{fSol}) of the restriction equation~(\ref{fRestriction})
reveals the seemingly local transformation~(\ref{localTrans}) to
be a nonlocal transformation of a dynamical variable,
\begin{equation}
\phi(t) \longrightarrow  \varphi(t) = \phi(t) - f_0
    - c_0 \int_{t_0}^{t} \! \frac{dt'}{\rho^2(t')} \, .
\label{NLtrans}
\end{equation}
Nonlocal transformations have to be approached with more care
when applied off-shell, compared to local transformations. In fact, the distinction 
between the two can have profound implications for operations that are usually taken for granted. 
One such operation is the step of implicitly disregarding the boundary
term in~(\ref{secondDS}), which would be justified for local transformations.
However, despite this contribution having the form of a boundary term,
\begin{equation}
{\tt (b.t.)}
    = \! \int_{t_a}^{t_b} \!\! dt \, \frac{d}{d t} 
    \biggl[ \frac{\rho^2 \dot{f}}{2} ( 2\varphi \!+\! f ) \biggr]
    \! =
    \biggl[ \frac{\rho^2 \dot{f}}{2} ( 2\varphi \!+\! f ) \biggr]_{t_a}^{t_b}
    \, ,
\end{equation}
plugging in the nonlocal solution~(\ref{fSol}) for the transformation 
function reveals it not to be a boundary term in the usual sense,
\begin{equation}
{\tt ( b.t.) } = 
    c_0 \bigl[ \varphi(t_b) - \varphi(t_a) \bigr]
    +
    \int_{t_a}^{t_b} \! dt \, \biggl[ \frac{ c_0^2}{2\rho^2(t)}
    \biggr] \, .
\label{bt}
\end{equation}
Rather, this would-be boundary term depends nonlocally on the dynamical
field~$\rho$, and thus contributes to the bulk action as
\begin{align}
\MoveEqLeft[2]
S[\phi,\rho] 
    \quad \longrightarrow \quad
    S[\varphi,\rho] 
\label{TransBulkAction}
\\
&
    =
    \int_{t_a}^{t_b} \! dt \, \biggl[
    \frac{\dot{\rho}^2}{2}
    + \frac{\rho^2 \dot{\varphi}^2}{2}
    - U(\rho)
    +
    \frac{ c_0^2}{2\rho^2}
    \biggr]
    +
    (\overline{\tt{b.t}})
    \, ,
\nonumber 
\end{align}
where now~$(\overline{\tt{b.t}})$ stands for the first term on the
right-hand side of~(\ref{bt}).
Thus, despite the superficial appearance, the transformation in~(\ref{localTrans}) 
with the restriction in~(\ref{fRestriction}) is {\it neither} a local 
transformation, {\it nor} a symmetry transformation, since both the
action and the equations of motion transform under it.

Having uncovered that the action transforms into~(\ref{TransBulkAction}),
it is no longer surprising that the equations of motion do not remain 
invariant. However, the equations of motion descending from applying 
the action principle to the action in~(\ref{TransBulkAction}),
\begin{equation}
\ddot{\rho} - \rho \dot{\varphi}^2 + U'(\rho) 
	= 
	- \frac{c_0^2}{\rho^3} \, ,
\qquad
\frac{d}{dt} \bigl( \rho^2 \dot{\varphi} \bigr) = 0 \, ,
\label{WrongRhoPhiEqs}
\end{equation}
still do not correspond to equations~(\ref{rehoEqTrans2})
and~(\ref{phiEqTrans2}), obtained by applying the transformation
directly to the original equations in~(\ref{EOMrho}) and~(\ref{EOMphi}).
This is true despite accounting for the nonlocal nature of the
would-be boundary term~(\ref{bt}),
which suggests the presence of
further obstructions with the way we had applied the 
transformation off-shell.
Resolving this issue requires us to first understand additional 
subtleties introduced by nonlocal 
transformations, which we consider in the following section.

\section{Nonlocal transformations}
\label{sec: Nonlocal transformations}

The case examined in the preceding section has taught us that
nonlocal transformations of the dynamical variables can generate
would-be boundary term contributions to the action, that are in actuality 
bulk contributions that have to be included in the action principle.
However, this is not the only subtle property of nonlocal 
transformations compared to the ubiquitous local transformations.
The very action principle might need modifications upon a nonlocal transformation: 
the variation of dynamical variables in general cannot
be taken to vanish at endpoints when deriving equations of 
motion.\footnote{We are grateful to Richard P.~Woodard for pointing this out.}
In this section, we examine a particularly simple example that 
demonstrates this feature.

Let us consider a system of two free non-interacting particles.
The dynamics of the system is encoded in the action,
\begin{equation}
S[X,Y] = \int_{t_a}^{t_b} \! dt \, 
    \biggl[
    \frac{\dot{X}^2}{2} + \frac{\dot{Y}^2}{2}
    \biggr]
    \, ,
\label{XYaction}
\end{equation}
and in the boundary conditions at the endpoints,
\begin{equation}
X(t_b) = x_b \, ,
\quad
Y(t_b) = y_b \, ,
\quad
X(t_a) = x_a \, ,
\quad
Y(t_a) = y_a \, .
\label{XYbcs}
\end{equation}
The standard action principle states that the true (on-shell) 
trajectories~$X(t)$ and~$Y(t)$ extremize the action, and thus 
the small variations~$\delta X(t)$ and~$\delta Y(t)$ do not
change the action,
\begin{align} \label{deltaS}
\MoveEqLeft[0.5]
\delta S[X,Y] 
    =
    S[X \!+\! \delta X , Y \!+\! \delta Y] - S[X,Y]
\nonumber \\
&   =
    \dot{X}(t_b) \delta X(t_b) 
    + 
    \dot{Y}(t_b) \delta Y(t_b)
    - 
    \dot{X}(t_a) \delta X(t_a) 
\\
&
    - 
    \dot{Y}(t_a) \delta Y(t_a)
    -\!
    \int_{t_a}^{t_b} \! dt \, \Bigl[ \ddot{X}(t) \delta X(t) 
        + \ddot{Y}(t) \delta Y(t) \Bigr]
    =
    0
    \, .\nonumber
\end{align}
The crucial part of the standard action principle is that the variations of the 
trajectories,~$\delta X(t)$ and~$\delta Y(t)$, are taken to vanish at endpoints
\begin{equation}
\delta X(t_b) = \delta Y(t_b) = \delta X(t_a) = \delta Y(t_b) = 0 \, .
\end{equation}
Only then does the middle line in~(\ref{deltaS}) vanish, and thus the equations of motion follow
\begin{equation}
\ddot{X} = 0 \, ,
\qquad
\ddot{Y} = 0 \, .
\label{XYeom}
\end{equation}
The solutions to these equations that respect the boundary 
conditions in~(\ref{XYbcs}) are then easily found
\begin{align}
&
X(t) = x_a + \frac{t \!-\! t_a}{t_b \!-\! t_a} (x_b \!-\! x_a) \, ,
\\
&
Y(t) = y_a + \frac{t \!-\! t_a}{t_b \!-\! t_a} (y_b \!-\! y_a) \, .
\end{align}

Let us now consider an invertible nonlocal transformation of the dynamical variables,
\begin{align}
&
X(t) \longrightarrow \mathcal{X}(t) = X(t) - c \int_{t_a}^{t} \! dt' \, Y(t') \, ,
\label{XYnonlocalTrans}
\\
&
Y(t) \longrightarrow \mathcal{Y}(t) = Y(t) \, ,
\end{align}
where~$c$ is some arbitrary constant. The equations of motion~(\ref{XYeom}) 
do transform, but remain local,
\begin{equation}
\ddot{\mathcal{X}} + c \dot{\mathcal{Y}} = 0 \, ,
\qquad
\ddot{\mathcal{Y}} = 0 \, .
\label{calXYcorrectEOM}
\end{equation}
The same holds for the action in~(\ref{XYaction}), where the integrand transforms, 
but remains local without any partial integration,
\begin{equation}
S[\mathcal{X},\mathcal{Y}] = \int_{t_a}^{t_b} \! dt \,
    \biggl[
    \frac{( \dot{\mathcal{X}} + c \mathcal{Y} )^2}{2}
    +
    \frac{\dot{\mathcal{Y}}^2}{2}
    \biggr]
    \, .
\end{equation}
Varying this action now gives
\begin{align}
\MoveEqLeft[2]
\delta S[\mathcal{X},\mathcal{Y}]
    =
    S[\mathcal{X} \!+\! \delta \mathcal{X} , \mathcal{Y} \!+\! \delta \mathcal{Y} ] 
    - 
    S[\mathcal{X},\mathcal{Y}]
\nonumber \\
={}&
    \bigl[ \dot{\mathcal{X}}(t_b) + c \mathcal{Y}(t_b) \bigr] \delta \mathcal{X}(t_b)
    +
    \dot{\mathcal{Y}}(t_b) \delta \mathcal{Y}(t_b)
\nonumber \\
&
    -
    \bigl[ \dot{\mathcal{X}}(t_a) + c \mathcal{Y}(t_a) \bigr] \delta \mathcal{X}(t_a)
    -
    \dot{\mathcal{Y}}(t_a) \delta \mathcal{Y}(t_a)
\label{calXYactionTrans} \\
&
    - \int_{t_a}^{t_b} \! dt \, 
    \Bigl\{
    ( \ddot{\mathcal{X}} + c \dot{\mathcal{Y}} ) \delta \mathcal{X}
    +
    \bigl[ \ddot{\mathcal{Y}} - c ( \dot{\mathcal{X}} + c \mathcal{Y} ) \bigr] \delta \mathcal{Y}
    \Bigr\}
    \, .
\nonumber
\end{align}
Naively we would set the variations at boundaries to vanish, and infer that the equations of motion descending from the transformed action are
\begin{equation}
\ddot{\mathcal{X}} + c \dot{\mathcal{Y}} = 0 \, ,
\qquad
\ddot{\mathcal{Y}} - c ( \dot{\mathcal{X}} + c \mathcal{Y} ) = 0 \, .
\label{calXYwrongEOM}
\end{equation}
These are, however, not the same equations obtained by redefining 
the equations of motion~(\ref{calXYcorrectEOM}). 
Here we encounter the same issue as by the end of 
Sec.~\ref{sec: Second example}, but this time without having to 
worry about partial integration in the transformed action.
This means that the explanation for the seeming discrepancy 
between equations~(\ref{calXYcorrectEOM}) and~(\ref{calXYwrongEOM})
should be sought for elsewhere.

The resolution of this seeming paradox is found in a closer examination of 
boundary conditions.
The nonlocal nature of the transformation in~(\ref{XYnonlocalTrans}) also
manifests itself in relations between boundary conditions for new and old variables,
%
\begin{align}
&
\mathcal{X}(t_b) = x_b -  c \int_{t_a}^{t_b} \! dt \, \mathcal{Y}(t) \, ,
\label{NLbc}
\\
&
\mathcal{Y}(t_b) = y_b \, ,
\qquad
\mathcal{X}(t_a) = x_a \, ,
\qquad
\mathcal{Y}(t_a) = y_a \, .
\label{XYbcs}
\end{align}
Here we notice that nonlocal dependence on the dynamical fields has sneaked into
the boundary condition~(\ref{NLbc}) after the transformation.
Neglecting this nonlocal dependence is what leads to the error 
in~(\ref{calXYactionTrans}), as the variations of new variables cannot 
be taken to vanish at endpoints. In fact, 
it follows from the transformation~(\ref{XYnonlocalTrans}) that the
variations at endpoints 
must be
\begin{align}
&
\delta \mathcal{X}(t_b) = - c \int_{t_a}^{t_b} \! \! dt \, \delta \mathcal{Y}(t) \, ,
\label{nonvanishingDX}
\\
&
\delta \mathcal{Y}(t_b)
    =
    \delta \mathcal{X}(t_a)
    =
    \delta \mathcal{Y}(t_a)
    =
    0
    \, .
\end{align}
Accounting for the non-vanishing endpoint variation~(\ref{nonvanishingDX}) 
in the variation of the action in~(\ref{calXYactionTrans}) then gives,
\begin{align}
\MoveEqLeft[0]
\delta S[\mathcal{X},\mathcal{Y}]
=
    - \int_{t_a}^{t_b} \! dt \, 
    \biggl\{
    \Bigl[ \ddot{\mathcal{X}}(t) + c \dot{\mathcal{Y}}(t) \Bigr] \delta \mathcal{X}(t)
\\
&
    +
    \Bigl[ \ddot{\mathcal{Y}}(t) - c \bigl[  \dot{\mathcal{X}}(t) + c \mathcal{Y}(t) \bigr]
    +
    c \bigl[ \dot{\mathcal{X}}(t_b) + c \mathcal{Y}(t_b)\bigr]  \Bigr] 
    \delta \mathcal{Y}(t)
    \biggr\}
    \, .
\nonumber 
\end{align}
Demanding that this entire variation vanishes
generates the following equations of motion,
\begin{align}
&
\ddot{\mathcal{X}}(t) + c \dot{\mathcal{Y}}(t) = 0 \, ,
\\
&
\ddot{\mathcal{Y}}(t) - c \bigl[  \dot{\mathcal{X}}(t) + c \mathcal{Y}(t) \bigr]
    +
    c \bigl[ \dot{\mathcal{X}}(t_b) + c \mathcal{Y}(t_b)\bigr]
    = 0 \, ,
\end{align}
{that are seemingly still not in the form of equations in~(\ref{calXYcorrectEOM}).
However, this discrepancy is only superficial, and, in fact, these equations 
indeed are equivalent to equation in~(\ref{calXYcorrectEOM}).}
This follows upon realizing that the first equation 
dictates~$\dot{\mathcal{X}}(t) \!+\! c\mathcal{Y}(t) \!=\! {\tt const.}$,
so that the two brackets in the second equation cancel.

This example teaches us that nonlocal field transformations can be implemented as actual transformations at the level of the action only if the nonlocalities they introduce into the boundary conditions are properly accounted for. If, on the other hand, such nonlocal dependencies in the boundary conditions are neglected, the relation between two sets of dynamical variables can no longer be considered a transformation. Rather, such prescription changes physics, as is manifested by the different equations of motion it produces. In conclusion, one should be very careful when employing nonlocal transformations,
as overlooking the details can change physics.

It should be noted that not all nonlocal transformations of dynamical 
variables lead to boundary variations that contribute to equations of motion.
For example, the Helmholtz decomposition of vector fields,
ubiquitous in electromagnetism~\cite{Griffiths:1492149}, and the
scalar-vector-tensor decomposition of cosmological 
perturbations~\cite{Lifshitz:1945du,Mukhanov:1990me} are 
such nonlocal field redefinitions that vanish at boundaries
and do not generate boundary variations. 
However, we ought to be aware
that nonlocal transformations {\it can} give rise to subtleties
described in this section. This is pertinent to restricted local transformations
considered in this work, where the restriction equation in general does
not have to allow for solutions describing nonlocal transformations vanishing
at the boundaries.

\section{Second example revisited}
\label{sec: Second example revisited}

In the preceding section, we demonstrated that nonlocal transformations have to be handled with due 
care at the level of the action. Having this in mind, 
we return here to the example of Sec.~\ref{sec: Second example} to explain the apparent paradox of its equations of motion after the transformation.

We have already revealed the transformation in~(\ref{localTrans}) 
to be nonlocal, generating nonlocal contributions to the bulk action 
via the would-be boundary terms. The full resulting action 
after the transformation reads
\begin{align}
S[\rho,\varphi] ={}& \int_{t_a}^{t_b}\! dt \, \biggl[
    \frac{\dot{\rho}^2}{2}
    + \frac{\rho^2 \dot{\varphi}^2}{2}
    - U(\rho)
    +
    \frac{c_0^2}{2\rho^2}
    \biggr]
\nonumber \\
&
    +
    c_0 \bigl[ \varphi(t_b) - \varphi(t_a) \bigr]
    \, .
\label{revisitedAction}
\end{align}
Implicit in the original action~(\ref{1Daction}) were the boundary conditions,
\begin{equation}
\rho(t_b) = \rho_b \, ,
\quad
\phi(t_b) = \phi_b \, ,
\quad
\rho(t_a) = \rho_a \, ,
\quad
\phi(t_a) = \phi_a \, .
\end{equation}
While boundary conditions for~$\rho(t)$ are not changed by the 
transformation~(\ref{localTrans}), boundary conditions for~$\varphi(t)$ pick up a 
nonlocal dependence on~$\rho(t)$,
\begin{align}
&
\varphi(t_b) = \phi_b - f_0 
    - c_0 \int_{t_0}^{t_b} \! \frac{dt}{\rho^2(t)} \, ,
\label{VarphiBoundary1}
\\
&
\varphi(t_a) = \phi_a - f_0
    - c_0 \int_{t_0}^{t_a} \! \frac{dt}{\rho^2(t)} \, .
\label{VarphiBoundary2}
\end{align}
This implies that boundary variations of~$\varphi(t)$ cannot be taken to vanish.
Rather, by varying~(\ref{VarphiBoundary1})
and~(\ref{VarphiBoundary2}) we find
\begin{equation}
\delta \varphi(t_b) = 2c_0 \!\! \int_{t_0}^{t_b} \!\!\! \frac{dt}{\rho^3(t)} \delta \rho(t) \, ,
\quad
\delta \varphi(t_a) = 2c_0 \!\! \int_{t_0}^{t_a} \!\!\! \frac{dt}{\rho^3(t)} \delta \rho(t) \, .
\label{BoundVar}
\end{equation}
Note that we cannot find a nontrivial solution to the restriction 
equation~(\ref{fRestriction}) that leads to both boundary variations above 
to vanish.
Taking~(\ref{BoundVar}) into account when varying the action
in~(\ref{revisitedAction}) 
then gives the equations of motion,
\begin{align}
&
\ddot{\rho} - \rho \dot{\varphi}^2 + U'(\rho)
    =
    \frac{c_0^2}{\rho^3}
    +
    \frac{2c_0}{\rho^3}
    \Bigl[
        \rho^2(t_b) \dot{\varphi}(t_b) \theta(t \!-\! t_0)
\label{RhoEqX}
\nonumber \\
&   \hspace{3cm}
        +
        \rho^2(t_a) \dot{\varphi}(t_a) \theta(t_0 \!-\! t)
        \Bigr]
        \, ,
\\
&
\frac{d}{dt} \bigl( \rho^2 \dot{\varphi} \bigr) = 0 \, .
\label{secEq}
\end{align}
We seem to have once more obtained equations different 
than~(\ref{rehoEqTrans2}) and~(\ref{phiEqTrans2}).
However, this time the difference is only superficial. The second 
equation~(\ref{secEq}) is a total derivative, which 
implies~$\rho^2(t_a) \dot{\varphi}(t_a) \!=\! 
\rho^2(t_b) \dot{\varphi}(t_b) \!=\! \rho^2 \dot{\varphi} $,
so that the right-hand
side of~(\ref{RhoEqX}) can, in fact, be rewritten to read
\begin{equation}
    \ddot{\rho} - \rho \dot{\varphi}^2 + U'(\rho)
    =
    \frac{c_0^2}{\rho^3}
    +
    \frac{2c_0\dot{\varphi}}{\rho}
    \, .
\end{equation}
This is now precisely Eq.~(\ref{rehoEqTrans2}), that is obtained by
applying the transformation~(\ref{NLtrans}) directly to the original
equations of motion.
Therefore, the seeming paradox from the end of
Sec.~\ref{sec: Second example}
disappears when nonlocalities in the boundary conditions are properly
accounted for. The fact that equations of motion are not left 
invariant under~(\ref{localTrans}) with the
restriction~(\ref{fRestriction}) is now explained:
that transformation is neither local, nor is it a symmetry 
transformation, {nor does it keep boundary
conditions local.}

\section{Third example: scale symmetry}
\label{sec: Third example}

Having understood the nuances of nonlocal transformations of dynamical
variables, we are now equipped to apply the lessons to systems exhibiting 
scale symmetry. The first example we consider is a one-dimensional
analogue of the pure~$R^2$ theory, that is supposed to be invariant under
restricted Weyl transformations~\cite{Edery:2014nha,Edery:2015wha}.
This system is given by the action
\begin{equation}
S[\sigma]
    =
    \int_{t_a}^{t_b} \! dt \, 
    \biggl(
    \frac{\ddot{\sigma} }{ \sigma }
    - \frac{ \dot{\sigma}^2}{2\sigma^2}
    \biggr)^{\!2} \, ,
\label{sigmaAction}
\end{equation}
together with the boundary conditions
\begin{equation}
\sigma(t_b) = \sigma_b \, ,
\quad
\dot{\sigma}(t_b) = \dot{\sigma}_b \, ,
\quad
\sigma(t_a) = \sigma_a \, ,
\quad
\dot{\sigma}(t_a) = \dot{\sigma}_a \, .
\label{sigmaBCs}
\end{equation}
The dynamical variable here is analogous to the metric in the~$R^2$
theory: the first term in parentheses in \eqref{sigmaAction} is the analogue of the
derivative of the Christoffel symbol, while the second term is the analogue
of the Christoffel symbol squared. 
{This is an imperfect analogy, but this simple theory
shares the essential properties of the pure~$R^2$
theory that are relevant for the points we are making in this work.}

Note that, because~(\ref{sigmaAction}) defines a higher derivative theory, we need to specify boundary conditions for both the dynamical variable and for its time derivative in~(\ref{sigmaBCs}). 
Consequently, the action principle assumes that variations of the 
dynamical variable and its time derivative vanish,
\begin{equation}
\delta \sigma(t_b) = \delta \sigma(t_a) = 0 \, ,
\qquad
\delta \dot{\sigma}(t_b) = \delta \dot{\sigma}(t_a) = 0 \, 
\end{equation}
and it produces the equation of motion,
\begin{equation}
\frac{d}{dt}
\biggl(
    \frac{d}{dt} - \frac{ \dot{\sigma} }{ \sigma } 
    \biggr)
    \biggl(
    \frac{\ddot{\sigma} }{ \sigma } - \frac{ \dot{\sigma}^2}{2\sigma^2}
    \biggr)
    =
    0
    \, ,
\label{sigmaEOM}
\end{equation}
where derivative operators act on everything to the right of them.
The action in~(\ref{sigmaAction}), and consequently the equation of motion~(\ref{sigmaEOM}), are invariant under global scale transformations~$\sigma(t) \! \to\! \omega^2 \sigma(t)$ for constant $\omega$. We again
follow the prescription from~\cite{Edery:2014nha,Edery:2015wha},
and promote this global symmetry to a 
tentative local restricted symmetry, given by the 
transformation rule
\begin{equation}
\sigma(t) \longrightarrow \Sigma(t) = \omega^{-2}(t) \sigma(t) \, ,
\label{sigmaTrans}
\end{equation}
where~$\omega(t)$ is supposed to satisfy an as-of-yet undetermined restriction condition so as to leave the action invariant. This restriction is best found by plugging in the transformation~(\ref{sigmaTrans}) into the action~(\ref{sigmaAction}), yielding
\begin{equation}
S[\sigma] \quad \longrightarrow \quad
S[\Sigma] =
    \int_{t_a}^{t_b} \! dt \, \biggl(
    \frac{\ddot{\Sigma} }{ \Sigma }
    - \frac{ \dot{\Sigma}^2}{2\Sigma^2}
    + \frac{2}{\omega} \mathcal{D} \omega
    \biggr)^{\!2}
    \, ,
\label{SigmaAction}
\end{equation}
where,
\begin{equation}
\mathcal{D} = 
    \biggl( \frac{d}{dt} + \frac{ \dot{\Sigma} }{ \Sigma } 
    \biggr) \frac{d}{dt} \, . \label{Ddef}
\end{equation}
It is now clear that requiring
\begin{equation}
\mathcal{D} \omega =  0 \, ,
\label{omegaRestriction}
\end{equation}
as a restriction condition, guarantees that the action remains invariant under the transformation in~(\ref{sigmaTrans}). 
The restriction condition~(\ref{omegaRestriction}) can be solved 
for~$\omega(t)$,
\begin{equation}
    \omega(t)
    =
    \omega_0
    +
    c_0 \int_{t_a}^{t}\frac{dt'}{\Sigma(t')}
    \, ,
\label{omegaSolution}
\end{equation}
showing that the transformation~(\ref{sigmaTrans})
with the restriction~(\ref{omegaRestriction}) is in fact a nonlocal transformation. 
Here~$\omega_0$ and~$c_0$ are constants of integration, the former one echoing the invariance under global scale transformations.

Given that the action remains invariant after such a transformation, the equation 
naively descending from the transformed action (without taking into account the nonlocalities in the boundary conditions) corresponds
to the one in~(\ref{sigmaEOM}), just with~$\sigma$ substituted by~$\Sigma$.
However, when the transformation in~(\ref{sigmaTrans}) is applied
to the equation of motion~(\ref{sigmaEOM}) directly, it results in
a transformed equation
\begin{equation}
\frac{d}{dt}
\biggl(
    \frac{d}{dt} - \frac{ \dot{\Sigma} }{ \Sigma } 
    \biggr)
    A
    =
    2 \dot{\omega} K
    \, ,
\label{SigmaEOM}
\end{equation}
where we have defined the combination on the left-hand side,
\begin{equation}
A(t) =
    \frac{\ddot{\Sigma} }{ \Sigma } 
    -
    \frac{ \dot{\Sigma}^2}{2\Sigma^2}
    \, ,
\label{Adef}
\end{equation}
and the source on the right-hand side as
\begin{equation}
K(t) = 
    \frac{1}{\omega}
    \biggl(
    \frac{d}{dt}
    -
    \frac{\dot{\Sigma}}{\Sigma}
    -
    \frac{\dot{\omega}}{\omega}
    \biggr)
    A
    =
    \Sigma
    \frac{d}{dt}
    \biggl(
    \frac{A}{\omega\Sigma}
    \biggr)
    \, .
\label{Kdef}
\end{equation}
We will see that this equation indeed does follow from the transformed action provided that the transformation is properly applied to the
boundary conditions as well.

Even though the action remains invariant under the transformation, the boundary conditions in~(\ref{sigmaBCs}) do not, due to the nonlocal nature of ~(\ref{sigmaTrans}). This also affects the variations at the boundaries, which read
\begin{align}
&
\delta \Sigma(t_b) = 2 \Sigma_b^2 \frac{\dot{\omega}_b}{\omega_b}
    \int_{t_a}^{t_b} \! dt \, \frac{ \delta \Sigma(t) }{\Sigma^2(t)}
\, , \quad
\delta \Sigma(t_a) = 0 \, ,
\label{deltaSigma}
\\
&
\delta \dot{\Sigma}(t_b) =
    2 \Sigma_b^2 \frac{\dot{\omega}_b}{\omega_b}
    \biggl( \frac{ \dot{\Sigma}_b}{\Sigma_b} 
        \!+\! \frac{\dot{\omega}_b}{\omega_b} \biggr)
    \int_{t_a}^{t_b} \! dt \, \frac{ \delta \Sigma(t) }{\Sigma^2(t)} \, ,
\quad
\delta \dot{\Sigma}(t_a) = 0 \, ,
\label{ddeltaSigma}
\end{align}
where, as usual, the subscripts~$a$ and~$b$ 
denote that the given quantity is evaluated at~$t_a$ or~$t_b$, respectively. Taking these non-vanishing boundary 
variations into account in the transformed action then leads to the following equation of motion,
\begin{equation}
\frac{d}{dt}
\biggl(
    \frac{d}{dt} - \frac{ \dot{\Sigma} }{ \Sigma } 
    \biggr)
    A
    =
    2 \dot{\omega} K_b
    \, ,
\label{SigmaEOM2}
\end{equation}
where~$K_b \!=\! K(t_b)$.
The above equation does not quite match the one obtained in~(\ref{SigmaEOM})  by directly transforming the original equation of motion. The difference is in the source on the right-hand side that in~(\ref{SigmaEOM}) is evaluated at the reference time of the equation,
while in~(\ref{SigmaEOM2}) it is evaluated at the final time~$t_b$.
However, this difference is again just superficial, as Eq.~(\ref{SigmaEOM2}) into Eq~(\ref{SigmaEOM}) can be shown to be the same equation, by appealing to conserved quantities encoded in the former.

Equation~(\ref{SigmaEOM2}) is already written as a total derivative with respect to the reference time~$t$, and can readily be integrated once, yielding
\begin{equation}
\omega K
    +
    \frac{\dot{\omega} A}{\omega}
    =
    2 \omega K_b
    +
    \Xi \, ,
\label{XiIntegrationConstant}
\end{equation}
where we have recognized the quantity defined in~(\ref{Kdef}),
and where~$\Xi$ is the constant of integration. 
Given that Eq.~(\ref{XiIntegrationConstant}) has to be satisfied
for all times, it is in particular satisfied at~$t\!=\!t_b$,
which determines the constant of integration in terms of
boundary conditions,
\begin{equation}
    \Xi
    =
    -
    \omega_b K_b
    +
    \frac{\dot{\omega}_b A_b}{\omega_b}
    \, .
\label{XiSolution}
\end{equation}
Equation~(\ref{XiIntegrationConstant}) can be written in a total
derivative form, by first multiplying it by~$1/\Sigma$, and then 
recalling that~$\dot{\omega}\!=\! c_0/\Sigma$. Upon that, it is readily integrated to yield,
\begin{equation}
    \frac{ A }{ \Sigma }
    =
    \frac{ \omega^2 }{c}
    K_b
    +
    \frac{\Xi \omega }{c}
    +
    \Theta
    \, ,
\label{ThetaIntegration}
\end{equation}
where~$\Theta$ is another constant of integration.
Exploiting again that the expression must be satisfied for all times, we evaluate it at~$t\!=\!t_b$ to find~$\Theta\!=\!0$, after 
applying~(\ref{XiSolution}). Now utilizing Eq.~(\ref{ThetaIntegration}) to eliminate~$A$ from~(\ref{XiIntegrationConstant}) gives
\begin{equation}
K_b = K \, .
\end{equation}
This is precisely the sought for result, that establishes the equivalence
of the equation of motion~(\ref{SigmaEOM2}), descendent from the 
transformed action, to the equation of motion~(\ref{SigmaEOM}),
obtained by transforming the original equations of motion.

Thus, we have proven that the transformation in~(\ref{sigmaTrans})
with the restriction in~(\ref{omegaRestriction}) (i) is a nonlocal
transformation, and (ii) is not a symmetry transformation despite
the naive transformation of the action~(\ref{SigmaAction}) suggesting so.

\section{Fourth example: more scale symmetry}
\label{sec: Fourth example}

We now consider the example of a globally scale invariant system
with two dynamical variables, that is a one-dimensional analogue
of the scalar-tensor theory in~(\ref{STaction}).
It is given by the action
\begin{equation}
 S[\sigma , \psi] 
    = 
    \int_{t_a}^{t_b} \! dt \,
    \biggl[ 
    \alpha \biggl( \frac{\ddot{\sigma}}{\sigma} 
        \!-\! \frac{\dot{\sigma}^2}{2\sigma^2} \biggr)^{\!2}
        +
        \frac{\sigma \dot{\psi}^2}{2}\biggr]
        \, 
\label{SigmaPsiAction}
\end{equation}
and by boundary conditions both at the lower bound,
\begin{equation}
\sigma(t_a) = \sigma_a \, ,
\qquad
\dot{\sigma}(t_a) = \dot{\sigma}_a \, ,
\qquad
\psi(t_a) = \psi_a \, ,
\label{LowerBoundary}
\end{equation}
and at the upper bound,
\begin{equation}
\sigma(t_b) = \sigma_b \, ,
\qquad
\dot{\sigma}(t_b) = \dot{\sigma}_b \, ,
\qquad
\psi(t_b) = \psi_b \, .
\label{UpperBoundary}
\end{equation}
The action principle produces two equations of motion,
\begin{align}
 &\alpha\frac{d}{dt} \biggl( \frac{d}{dt} \!-\! \frac{\dot{\sigma}}{\sigma} \biggr)
    \biggl( \frac{ \ddot{\sigma} }{ \sigma } 
        - \frac{\dot{\sigma}^2 }{ 2\sigma^2 } \biggr)
    + \frac{ \sigma \dot{\psi}^2 }{4}
    =
    0
    \, ,
\label{EOMSigma}
\\
&  \frac{d}{dt} \bigl( \sigma \dot{\psi} \bigr) = 0 \, .
\end{align}
Both the action and the equations of motion are invariant under a global 
rescaling of dynamical variables,~$\sigma(t) \!\to\! \omega^2 \sigma(t)$
and~$\psi(t) \!\to\! \psi(t)/\omega$. We once more examine the proposal to 
promote this global symmetry to a restricted local one,
\begin{align}
\sigma(t) \longrightarrow{}& 
	\Sigma(t)
	= 
	\omega^{-2} (t) \sigma(t)
	\, ,
\label{SigmaPsiTrans1}
\\
\psi(t) \longrightarrow{}&
	\Psi(t)
	=
	\omega(t) \psi(t)
	\, ,
\label{SigmaPsiTrans2}
\end{align}
with the restriction that the transformation function satisfies,
\begin{equation}
\mathcal{D} \omega = 0 \, ,
\label{omCond}
\end{equation}
where~$\mathcal{D}$ was defined in~(\ref{Ddef}). 
This guarantees that the~$\Psi(t)$-independent part of the 
action~(\ref{SigmaPsiAction}) is invariant under the transformation,
while the remaining part is invariant up to boundary terms upon
partial integration,
\begin{align}
S[\sigma,\psi]
    \longrightarrow
    S[\Sigma,\Psi]
    ={}&
    \int_{t_a}^{t_b} \! dt \,
    \biggl[ 
    \alpha \biggl( \frac{\ddot{\Sigma}}{\Sigma} 
        \!-\! \frac{\dot{\Sigma}^2}{2\Sigma^2} \biggr)^{\!2}
    +
    \frac{\Sigma \dot{\Psi}^2 }{2}
    \biggr]
\nonumber \\
&
    -
    \frac{ \Psi_b^2 \Sigma_b \dot{\omega}_b }{2\omega_b}
    +
    \frac{ \Psi_a^2 \Sigma_a \dot{\omega}_a }{2\omega_a}
    \, .
\label{SigmaPsiAction}
\end{align}
The restriction condition~(\ref{omCond}) dictates that the
transformation function~(\ref{omegaSolution}) 
is a nonlocal function of one of the dynamical variables. Thus, the would-be
boundary terms in the transformed action above cannot be 
neglected. Their dependence on~$\omega_b\!=\!\omega(t_b)$ 
and~$\omega_a\!=\! \omega(t_a)$ implies a nonlocal dependence on~$\Sigma(t)$, though 
this dependence is such that it cannot be
written in terms of an integral over a local Lagrangian. Thus,
the transformation in~(\ref{SigmaPsiTrans1}) and~(\ref{SigmaPsiTrans2}) 
is not even naively a symmetry of the action.

Inferring the equations of motion encoded by the action 
in~(\ref{SigmaPsiAction}) requires us to recognize that the boundary
conditions in~(\ref{LowerBoundary}) and~(\ref{UpperBoundary}) 
also transform in a manner that introduces nonlocal dependence in them.
This further influences the variations employed in the action principle,
that can no longer be taken to vanish at endpoints. For the~$\Sigma(t)$
variable these variations are given in~(\ref{deltaSigma})
and~(\ref{ddeltaSigma}), while for the~$\Psi(t)$
variable they are,
\begin{equation}
\delta \Psi(t_b)
    =
    - 
    \frac{\Psi_b}{\omega_b}
    \int_{t_a}^{t_b}\! dt \, \frac{ \dot{\omega}(t) }{ \Sigma(t) } 
    \delta \Sigma(t)
    \, ,
\quad
\delta \Psi(t_a) 
    =
    0
    \, .
\end{equation}
Taking these into account, the action principle applied to 
action~(\ref{SigmaPsiAction}) then produces the following equations
of motion,
\begin{align}
&
    \alpha
    \frac{d}{dt}
    \biggl( \frac{d}{dt} \!-\! \frac{ \dot{\Sigma} }{ \Sigma } \biggr) A
    +
    \frac{\Sigma \dot{\Psi}^2}{4}
    =
    \dot{\omega} 
    \bigl(
    2 \alpha K_b
    +
    M_b
    \bigr)
    \, ,
\label{SigmaPsiEq1}
\\
&
    \frac{d}{dt} \bigl( \Sigma \dot{\Psi} \bigr) = 0 \, ,
\label{SigmaPsiEq2}
\end{align}
where~$A(t)$ and~$K(t)$ were defined in~(\ref{Adef})
and~(\ref{Kdef}), and where we introduced another definition,
\begin{equation}
M(t) = 
    \frac{\Sigma\Psi\dot \Psi}{2\omega}
    -
    \frac{c\Psi^2}{4\omega^2}
    =
    \frac{\Sigma}{4}
    \frac{d}{dt}
    \Bigl(
    \frac{\Psi^2 }{\omega}
    \Bigr)
    \, ,
\end{equation}
with~$M_b \!=\! M(t_b)$.

The right-hand side of the first equation in~(\ref{SigmaPsiEq1})
can be rewritten in a similar manner, compared to the previous section. We first integrate the second equation~(\ref{SigmaPsiEq2}),
\begin{equation}
\Sigma \dot{\Psi} = \Gamma \, ,
\end{equation}
where~$\Gamma$ is a constant of integration. We then use the result
to rewrite the second term on the left-hand side in Eq.~(\ref{SigmaPsiEq1})
\begin{equation}
    \alpha
    \frac{d}{dt}
    \biggl( \frac{d}{dt} \!-\! \frac{ \dot{\Sigma} }{ \Sigma } \biggr) A
    +
    \frac{\Gamma \dot{\Psi} }{4}
    =
    \dot{\omega} 
    \bigl(
    2 \alpha K_b
    +
    M_b
    \bigr)
    \, .
\end{equation}
This equation is now in a total derivative form, and can be readily 
integrated to give,
\begin{equation}
    \alpha 
    \Bigl( \omega
    K
    +
    \frac{ \dot{\omega} A}{\omega}
    \Bigr)
    +
    \frac{\Gamma \Psi }{4}
    =
    \omega
    \bigl(
    2 \alpha K_b
    +
    M_b
    \bigr)
    +
    \Xi
    \, ,
\label{OnceIntegrated}
\end{equation}
where~$\Xi$ is a constant of integration. This expression has to be valid
at all times, so we may evaluate it at~$t\!=\!t_b$ to express the
constant of integration in terms of boundary conditions,
\begin{equation}
    \Xi
    =
    \alpha 
    \Bigl( 
    - \omega_b
    K_b
    +
    \frac{ \dot{\omega}_b A_b}{\omega_b}
    \Bigr)
    +
    \frac{\Gamma \Psi_b }{4}
    -
    \omega_b M_b
    \, .
\end{equation}
Equation~(\ref{OnceIntegrated}) can be integrated once more. 
This can be accomplished by first multiplying 
expression~(\ref{OnceIntegrated}) by~$1/\Sigma$, 
\begin{equation}
    \alpha 
    \frac{d}{dt}
    \Bigl(
    \frac{A}{\Sigma}
    \Bigr)
    +
    \frac{\Gamma \Psi }{4 \Sigma}
    =
    \frac{ \omega }{ \Sigma }
    \bigl(
    2 \alpha K_b
    +
    M_b
    \bigr)
    +
    \frac{ \Xi }{ \Sigma }
    \, ,
\end{equation}
and then by recognizing that~$\Gamma\!=\! \Sigma \dot{\Psi}$ 
and~$1/\Sigma \!=\! \dot{\omega}/c$, so that the integral is
\begin{equation}
    \frac{\alpha A}{\Sigma}
    +
    \frac{ \Psi^2 }{8}
    =
    \frac{ \omega^2 }{ 2c }
    \bigl(
    2 \alpha K_b
    +
    M_b
    \bigr)
    +
    \frac{ \Xi }{ c } \omega
    +
    \Theta
    \, ,
\label{TwiceIntegrated}
\end{equation}
where~$\Theta$ is a constant of integration. Evaluating this expression
at~$t\!=\!t_b$ reveals that~$\Theta\!=\!0$. Then 
using~(\ref{TwiceIntegrated}) in~(\ref{OnceIntegrated})
to remove~$A(t)$, together with the expressions for constants of 
integration,
gives that the right-hand side of Eq.~(\ref{SigmaPsiEq1})
can be substituted by the time-dependent version,
\begin{equation}
    2 \alpha K_b
    +
    M_b
    =
    2\alpha K 
    +
    M
    \, .
\end{equation}
This is precisely the equation that follows from transforming the 
original equation of motion~(\ref{EOMSigma}), confirming that everything
works out consistently provided that the nonlocalities from the
would-be boundary terms in the action, and from the boundary variations
are properly taken into account. Therefore, the transformation 
in~(\ref{SigmaPsiTrans1}) and~(\ref{SigmaPsiTrans2})
with the restriction in~(\ref{omCond})
is neither a local transformation, nor a symmetry transformation
of the system given in~(\ref{SigmaPsiAction}).

\section{Scalar-tensor theory}
\label{sec: Scalar-tensor theory}

The examples considered in Secs.~\ref{sec: Second example}--\ref{sec: Fourth example} 
have taught us that ``restricted transformations'' are typically nonlocal.
At the level of the action neither boundary terms nor boundary variations can be taken lightly
when nonlocal transformations are concerned, and in general they both contribute to the equations
of motion. Here we apply these lessons to the case of the scalar-tensor theory, considered in the 
introductory section.

The conformal factor~$\Omega$ of the ``restricted Weyl 
transformation'' in~(\ref{restrictedWeylTransMetric}) and~(\ref{restrictedWeylTrans}) 
has to satisfy the covariant Klein-Gordon equation~(\ref{WeylRestriction})
where the d'Alembertian operator depends on the transformed metric.
This introduces a nonlocal dependence on the metric in the conformal factor.
An immediate consequence of this is seen for the boundary
term that arises from the scalar kinetic term upon applying the 
transformation~\cite{Edery:2014nha},
\begin{equation}
\sqrt{-g} \,
    \nabla^\mu \phi \nabla_\mu \phi
    \longrightarrow
    \sqrt{-g'}
    \biggl[ 
    \nabla'^\mu \phi' \nabla'_\mu \phi'
    \!-\!
    \nabla'^\mu \Bigl( \phi'^2 \frac{\partial_\mu \Omega}{\Omega} \Bigr)
    \biggr]
    \, .
\end{equation}
This would-be boundary term cannot be discarded because the conformal factor~$\Omega$
depends functionally on the transformed metric, and we have to vary this dependence
when deriving the equations of motion from the transformed action. 
However, working out the explicit functional dependence of~$\Omega$ on the metric is rather 
challenging. This also impedes us from working out 
explicitly the boundary variations of the fields upon applying the 
transformation. Nevertheless, illustrating the
subtle issues explicitly can be efficiently accomplished at the level of perturbations.

We consider nonlinear perturbations of the scalar-tensor theory in~(\ref{STaction})
around Minkowski space.\footnote{
There are subtleties with strong coupling for linear perturbations of the pure~$R^2$ theory 
around Minkowski space~\cite{Hell:2023mph,Alvarez-Gaume:2015rwa,Golovnev:2023zen,Karananas:2024hoh},
that are avoided by considering nonlinear perturbations as we do here.}
To this end, we expand the metric in small fluctuations,
\begin{equation}
g_{\mu\nu} = \eta_{\mu\nu} + \kappa h_{\mu\nu} \, ,
\end{equation}
while the scalar field  is treated as a perturbation itself. 
For our purposes it is sufficient to consider the cubic action for perturbations,
\begin{align}
\MoveEqLeft[0.5]
S[h_{\mu\nu},\phi] 
    =
    \int_{V} \! d^{4\!}x \, 
    \biggl[
    \alpha \kappa^2 \bigl( D^{\mu\nu} h_{\mu\nu} \bigr)^{\!2}
    -
    \frac{1}{2} (\partial^\mu\phi ) ( \partial_\mu \phi )
\nonumber \\
&
    +
    2 \alpha
    \kappa^3
    \Bigl( 
    {\rm T}_1^{\mu\nu\rho\sigma\omega\lambda}
    h_{\mu\nu} \partial_\omega \partial_\lambda h_{\rho\sigma}
    +
    {\rm T}_2^{\mu\nu\rho\sigma\omega\lambda}
    \partial_\omega h_{\mu\nu} \partial_\lambda h_{\rho\sigma}
    \Bigr)
\nonumber \\
&
    \times \!
    D^{\alpha\beta} h_{\alpha\beta}
    +
    \frac{\kappa }{4}
    \bigl( 2 h^{\mu\nu} \!-\! \eta^{\mu\nu} h \bigr)
        \partial_\mu\phi \, \partial_\nu \phi
    -
    \frac{\xi \kappa }{2} \phi^2 D^{\mu\nu} h_{\mu\nu}
    \biggr]
    \, ,
\label{PertAction}
\end{align}
where~$V$ is the integration volume, 
with boundary~$\partial V$
on which boundary conditions are specified.
Here we introduced a shorthand notation for 
derivatives,~$D^{\mu\nu} \!=\! \partial^\mu \partial^\nu 
    \!-\! \eta^{\mu\nu} \partial^2$, and
two tensor structures
\begin{align}
T_1^{\mu\nu\rho\sigma\omega\lambda}
    ={}&
    \eta^{\mu(\omega} \eta^{\lambda)\nu} \eta^{\rho\sigma}
    +
    \eta^{\mu(\rho} \eta^{\sigma)\nu} \eta^{\omega\lambda}
    +
    \frac{1}{4} \eta^{\mu\nu} \eta^{\rho(\omega} \eta^{\lambda)\sigma}
\nonumber \\
&
    -
    2 \eta^{\omega) (\mu} \eta^{\nu) (\rho} \eta^{\sigma) (\lambda}
    -
    \frac{1}{4} \eta^{\mu\nu} \eta^{\omega\lambda} \eta^{\rho\sigma}
    \, ,
\\
T_2^{\mu\nu\rho\sigma\omega\lambda}
    ={}&
    \frac{1}{2} \eta^{\omega(\mu} \eta^{\nu)\lambda} \eta^{\rho\sigma}
    \!+\!
    \frac{1}{2} \eta^{\omega(\rho} \eta^{\sigma)\lambda} \eta^{\mu\nu}
    \!-\!
    \frac{1}{4} \eta^{\mu\nu} \eta^{\rho\sigma} \eta^{\omega\lambda}
\nonumber \\
&   \hspace{-1.8cm}
    +
    \frac{3}{4}
    \eta^{\mu(\rho} \eta^{\sigma)\nu} \eta^{\omega\lambda}
    \!-\!
    \eta^{\omega(\mu} \eta^{\nu)(\sigma} \eta^{\rho)\lambda}
    \!-\!
    \frac{1}{2}
    \eta^{\omega (\rho} \eta^{\sigma) (\mu} \eta^{\nu) \lambda}
    \, .
\end{align}
The scalar and tensor equations of motion descend from the variational
principle applied to the action in~(\ref{PertAction}),
with vanishing boundary variations,
\begin{equation}
\delta h_{\mu\nu}(x) \big|_{\partial V} \! = 0 \, ,
\quad 
\partial_\rho \delta h_{\mu\nu}(x) \big|_{\partial V} \! = 0 \, ,
\quad
\delta \phi(x) \big|_{\partial V} \! = 0 \, .
\label{VanishingBoundary}
\end{equation}
We write the equations of motion as,
\begin{equation}
\mathcal{F}[h_{\rho\sigma},\phi] = 0 \, ,
\qquad
\mathcal{E}_{\mu\nu}[h_{\rho\sigma},\phi] = 0 \, ,
\label{STeoms}
\end{equation}
where we introduced two functionals, the scalar one,
\begin{equation}
\mathcal{F}[h_{\rho\sigma},\phi]
    =
    \partial^2 \phi
    -
    \frac{\kappa}{2} \partial_\mu 
    \bigl( 
    2 h^{\mu\nu} \partial_\nu \phi
    -
    h \partial^\mu \phi
    \bigr)
    -
    \xi \kappa \phi D^{\alpha\beta} h_{\alpha\beta}
    \, ,
\end{equation}
and the tensor one,
\begin{align}
\MoveEqLeft[0.5]
\mathcal{E}_{\mu\nu}[h_{\rho\sigma},\phi]
    =
    2 \alpha \kappa^2 \biggl\{
    D^{\mu\nu}
    \bigl( 
        \partial^\alpha \partial^\beta h_{\alpha\beta} 
        \!-\! 
        \partial^2 h 
        \bigr)
    +
    \kappa D^{\mu\nu}
\nonumber \\
&
    \times\!
    \Bigl[
    {\rm T}_1^{\alpha\beta\rho\sigma\omega\lambda}
    h_{\alpha\beta} \partial_\omega \partial_\lambda h_{\rho\sigma}
    +
    {\rm T}_2^{\alpha\beta\rho\sigma\omega\lambda}
    \partial_\omega h_{\alpha\beta} \partial_\lambda h_{\rho\sigma}
    \Bigr]
\nonumber \\
&
    +
    \kappa
    {\rm T}_1^{\mu\nu\rho\sigma\omega\lambda}
    \bigl( D^{\alpha\beta} h_{\alpha\beta} \bigr)
    \partial_\omega \partial_\lambda h_{\rho\sigma}
    +
    \kappa
    {\rm T}_1^{\rho\sigma\mu\nu\omega\lambda}
    \partial_\omega \partial_\lambda
\nonumber \\
&
    \!\times\!
    \bigl( h_{\rho\sigma}  D^{\alpha\beta} h_{\alpha\beta} \bigr)
    -
    2
    \kappa
    {\rm T}_2^{\mu\nu\rho\sigma\omega\lambda}
    \partial_\omega 
    \Bigl[
    \bigl( D^{\alpha\beta} h_{\alpha\beta} \bigr)
    \partial_\lambda h_{\rho\sigma}
    \Bigr]
    \biggr\}
\nonumber \\
&
    +
    \frac{\kappa}{2 }
    \Bigl[
    2 ( \partial^\mu \! \phi ) ( \partial^\nu \!\phi )
        \!-\! 
        \eta^{\mu\nu}
            (\partial^\rho \! \phi ) ( \partial_\sigma \phi )
    \!-\!
    2\xi D^{\mu\nu} \phi^2
    \Bigr]
    \, .
\end{align}

Let us now consider the perturbative version of the transformation
in~(\ref{restrictedWeylTrans}). The  function $\Omega(x)$ is expanded as
\begin{equation}
\Omega = 1 + \kappa \Omega_1 + \kappa^2 \Omega_2 \, ,
\end{equation}
so that the transformations reads,
\begin{align}
h_{\mu\nu}
    \longrightarrow{}&
    \gamma_{\mu\nu}
    =
    h_{\mu\nu}
    -
    2 \Omega_1 \eta_{\mu\nu}
\nonumber \\
&
    -
    \kappa \Bigl(
    2 \Omega_1 h_{\mu\nu}
    -
    3\Omega_1^2 \eta_{\mu\nu}
    + 
    2 \Omega_2 \eta_{\mu\nu}
    \Bigr)
    \, ,
\label{hTrans}
\\
\phi \longrightarrow{}& \varphi =
     \phi + \kappa \Omega_1 \phi 
     \, .
\label{phiTrans}
\end{align}
The restriction condition~(\ref{WeylRestriction}) then breaks up into
conditions at different perturbative orders. The first order condition
is the flat space Klein-Gordon equation,
\begin{equation}
\partial^2 \Omega_1 = 0 \, ,
\label{Omega1KG}
\end{equation}
while the second order condition is the sourced Klein-Gordon equation,
\begin{equation}
\partial^2 \Omega_2 
    =
    \gamma^{\mu\nu} \partial_\mu \partial_\nu \Omega_1
    +
    \frac{1}{2} \bigl( 2 \partial_\mu \gamma^{\mu\nu} 
        \!-\! \partial^\nu \gamma \bigr)
        \partial_\nu \Omega_1
    \, .
\label{Omega2eq}
\end{equation}
The equations of motion~(\ref{STeoms}) are not both invariant under ~(\ref{hTrans}) and~(\ref{phiTrans}). 
While the scalar equation does not transform,
\begin{equation}
\mathcal{F} [ \gamma_{\rho\sigma} , \varphi ] = 0 \, ,
\end{equation}
the tensor one does,
\begin{equation}
\mathcal{E}_{\mu\nu} [ \gamma_{\rho\sigma} , \varphi ]
    =
    6 \alpha \kappa^2
    \bigl( 2 \eta^{\rho(\mu} \eta^{\nu)\sigma}
        \!-\! \eta^{\mu\nu} \eta^{\rho\sigma} \bigr)
    \partial_\rho \Omega_1
    \partial_\sigma D^{\alpha\beta} \gamma_{\alpha\beta}
    \, .
\label{TensorEqTransformed}
\end{equation}
These  equations can be shown to correspond to a perturbative expansion of the
transformed equations from Sec. \ref{sec:IntroductionAndMotivation}.
We should note that, at the quadratic order in perturbations, the transformed equations of motion are still local in the dynamical variables, since
only~$\Omega_1$, that is a solution of the free Klein-Gordon 
equation~(\ref{Omega1KG}), appears there. At higher orders,~$\Omega_2$
will appear in the equations, making them nonlocal. The reason is
that~$\Omega_2$ itself depends nonlocally on the dynamical variables.
This is seen by solving Eq.~(\ref{Omega2eq}) using
the Green's function method,
\begin{align}
\Omega_2(x) ={}& 
    \int_V\! d^{4\!}x' \, G(x;x') 
    \biggl\{
    \gamma^{\mu\nu}(x') \partial'_\mu \partial'_\nu \Omega_1(x')
\nonumber \\
&
    +
    \Bigl[ \partial'_\mu \gamma^{\mu\nu}(x')
    \!-\! \tfrac{1}{2} \partial'^\nu \gamma(x') \Bigr]
    \partial'_\nu \Omega_1(x') 
    \biggr\}
    \, ,
\label{Omega2solution}
\end{align}
where the Green's function satisfies,
\begin{equation}
\partial^2 G(x;x') = \delta^4(x \!-\! x') \, .
\label{GreenEq}
\end{equation}
A complete specification of the Green's function includes fixing the 
boundary conditions. However, whichever one we choose is immaterial for the points we intend to demonstrate, as will become clear by the end of this section.

Next we turn to applying the transformations~(\ref{hTrans})
and~(\ref{phiTrans}) directly to the action,
which turns out to be invariant up to a boundary term,
\begin{align}
\MoveEqLeft[1]
S[h_{\mu\nu}, \phi] 
\quad \longrightarrow \quad
S[\gamma_{\mu\nu}, \varphi]
    =
    \int_{V}\!\! d^{4\!}x \, 
    \biggl[
    \alpha \kappa^2 \bigl( D^{\mu\nu} \gamma_{\mu\nu} \bigr)^{\!2}
\nonumber \\
&
    -
    \frac{1}{2} (\partial^\mu\varphi ) ( \partial_\mu \varphi )
    +
    \frac{\kappa}{2} \partial^\mu ( \varphi^2 \partial_\mu \Omega_1 )
    +
    2 \alpha
    \kappa^3
    \Bigl( 
    {\rm T}_1^{\mu\nu\rho\sigma\omega\lambda}
\nonumber \\
&
    \times \!
    \gamma_{\mu\nu} \partial_\omega \partial_\lambda \gamma_{\rho\sigma}
    +
    {\rm T}_2^{\mu\nu\rho\sigma\omega\lambda}
    \partial_\omega \gamma_{\mu\nu} \partial_\lambda \gamma_{\rho\sigma}
    \Bigr)
    D^{\alpha\beta} \gamma_{\alpha\beta}
\nonumber \\
&
    +
    \frac{\kappa }{4}
    \bigl( 2 \gamma^{\mu\nu} \!-\! \eta^{\mu\nu} \gamma \bigr)
        \partial_\mu\varphi \, \partial_\nu \varphi
    -
    \frac{\xi \kappa }{2} \varphi^2 D^{\mu\nu} \gamma_{\mu\nu}
    \biggr]
    \, ,
\label{PertAction2}
\end{align}
just like the transformed action from Sec.~\ref{sec:IntroductionAndMotivation}.
This boundary term actually does not contribute to equations of motion 
at the perturbative order are working at, but it will contribute at higher 
orders. Thus, approaching the transformed action naively, we might
conclude that the equations of motion must be invariant as well, which 
we know is not true. The resolution of this seeming paradox lies once
more in carefully considering the non-vanishing boundary variations 
introduced by the nonlocal transformation. The transformed scalar 
still retains vanishing boundary conditions,
\begin{equation}
\delta \varphi(x) \big|_{\partial V} \!= 0 \, ,
\end{equation}
but the transformed metric does not,
\begin{align}
\delta \gamma_{\mu\nu}(x) \big|_{\partial V} \!
    ={}& 
    - 2 \kappa \eta_{\mu\nu} \delta \Omega_2(x) \big|_{\partial V}
    \, ,
\label{gammaVar}
\\
\partial_\rho \delta \gamma_{\mu\nu}(x) \big|_{\partial V} \! 
    ={}&
    - 2 \kappa \eta_{\mu\nu} \partial_\rho \delta \Omega_2(x) \big|_{\partial V}
    \, ,
\label{delgammaVar}
\end{align}
on account of~$\Omega_2$ in~(\ref{Omega2solution}) depending
nonlocally on the transformed metric. These non-vanishing boundary 
variations will be important in the would-be boundary terms generated
by varying the action. Note that~(\ref{gammaVar}) and~(\ref{delgammaVar})
are of quadratic order in perturbations, and can only contribute to the
variation of the quadratic part of the action. The relevant would-be
boundary terms are
\begin{align}
\MoveEqLeft[1]
\delta S[\gamma_{\mu\nu}, \varphi] \, \Big|_{\tt b.t.}
    =
    2 \alpha \kappa^2 
    \bigl( \eta^{\rho(\mu} \eta^{\nu)\sigma}
        \!-\! \eta^{\mu\nu} \eta^{\rho\sigma}\bigr)
\\
&
    \times \!
    \int_V \! d^{4\!} x \, 
    \partial_\rho \Bigl[
    \bigl( D^{\alpha\beta} \gamma_{\alpha\beta} \bigr)
        \partial_{\sigma} \delta \gamma_{\mu\nu}
        \!-\!
        \partial_\sigma \bigl( D^{\alpha\beta} \gamma_{\alpha\beta} \bigr)
        \delta \gamma_{\mu\nu}
        \Bigr]
        \, .
\nonumber 
\end{align}
Plugging in the boundary variations~(\ref{gammaVar}) and~(\ref{delgammaVar}),
gives,
\begin{align}
\delta S[\gamma_{\mu\nu}, \varphi] \, \Big|_{\tt b.t.}
    ={}&
    12 \alpha \kappa^3
    \int_V \! d^{4\!} x \, 
    \partial^\rho \Bigl[
    \bigl( D^{\alpha\beta} \gamma_{\alpha\beta} \bigr)
        \partial_\rho \delta \Omega_2
\nonumber \\
&   \hspace{0.9cm}
        \!-\!
        \partial_\rho \bigl( D^{\alpha\beta} \gamma_{\alpha\beta} \bigr)
        \delta \Omega_2
        \Bigr]
        \, 
\end{align}
and then using the solution~(\ref{Omega2solution}) 
for~$\Omega_2$  gives,
\begin{align}
\MoveEqLeft[1.5]
\delta S[\gamma_{\mu\nu}, \varphi] \, \Big|_{\tt b.t.}
    \! =
    12 \alpha \kappa^3
    \int_V\! d^{4\!}x' \, 
    \biggl\{
    \delta \gamma^{\mu\nu}(x') \partial'_\mu \partial'_\nu \Omega_1(x')
\nonumber \\
&   \hspace{1cm}
    +
    \Bigl[ \partial'_\mu \delta\gamma^{\mu\nu}(x')
    \!-\! \tfrac{1}{2} \partial'^\nu \delta\gamma(x') \Bigr]
    \partial'_\nu \Omega_1(x')
    \biggr\}
\nonumber \\
&
    \times\!
    \int_V \! d^{4\!} x \, 
    \partial^\rho
    \Bigl\{
    \bigl[ D^{\alpha\beta} \gamma_{\alpha\beta}(x) \bigr]
        \partial_\rho G(x;x') 
\nonumber \\
&   \hspace{2.2cm}
        -
        \partial_\rho 
        \bigl[ D^{\alpha\beta} \gamma_{\alpha\beta}(x) \bigr]
        G(x;x') 
        \Bigr\}
        \, .
\end{align}
This expression is cubic in perturbations, so we can partially integrate
derivatives away from the metric without generating further boundary
contributions. This finally produces the right-hand side of the 
tensor equation of motion,
\begin{align}
\MoveEqLeft[2]
\mathcal{E}_{\mu\nu}[\gamma_{\rho\sigma},\varphi]
    =
    6 \alpha \kappa^3
        \bigl( 2\delta_\mu^\rho \delta_\nu^\sigma 
            \!-\! \eta_{\mu\nu} \eta^{\rho\sigma} \bigr)
        \partial_\rho \Omega_1 
\nonumber \\
&
    \times\! 
    \partial_\sigma \int_V \! d^{4\!} x' \, 
    \partial'^\lambda
    \Bigl\{
    \bigl[ D'^{\alpha\beta} \gamma_{\alpha\beta}(x') \bigr]
        \partial'_\lambda G(x';x) 
\nonumber \\
&   \hspace{2.1cm}
        -
        \partial'_\lambda 
        \bigl[ D'^{\alpha\beta} \gamma_{\alpha\beta}(x') \bigr]
        G(x';x) 
        \Bigr\}
        \, .
\label{TensorEq3}
\end{align}
Here we once more encounter an equation that seemingly does not correspond
to the one in~(\ref{TensorEqTransformed}) that we are supposed to obtain.
This apparent discrepancy is again just superficial. From the trace of the 
leading order equation,
\begin{equation}
\partial^2 \bigl( D^{\alpha\beta} \gamma_{\alpha\beta} \bigr) 
    = 
    \mathcal{O}(\text{\tt quadratic order}) \, ,
\end{equation}
and the Green's function equation~(\ref{GreenEq}) we find that
\begin{align}
\MoveEqLeft[2]
\int_V \! d^{4\!} x' \, 
    \partial'^\lambda \Bigl\{
    \bigl[ D'^{\alpha\beta} \gamma_{\alpha\beta}(x') \bigr]
        \partial'_\lambda G(x';x) 
\nonumber \\
&
        -
        \partial'_\lambda 
        \bigl[ D'^{\alpha\beta} \gamma_{\alpha\beta}(x') \bigr]
        G(x';x)
        \Bigr\}
    =
    D^{\alpha\beta} \gamma_{\alpha\beta}(x)
    \, .
\label{GreenId}
\end{align}
This identity is independent of the particular boundary conditions chosen for 
the Green's function.
Plugging  this result into the right-hand side of~(\ref{TensorEq3})
then reproduces precisely the transformed tensor equation~(\ref{TensorEqTransformed}).

\section{Discussion}
\label{sec: Discussion}

``Restricted Weyl symmetry'' is a proposed class of symmetry transformations 
somewhere in between global scale transformations and local Weyl 
transformations. 
They take the form of local conformal rescalings of fields, 
with the additional requirement that the conformal factor satisfies the covariant 
Klein-Gordon equation.
Such transformations indeed leave the action for a 
certain class of scalar-tensor theories formally invariant up to boundary 
terms. However, it seems to have gone unnoticed,
in the literature thus far~\cite{Edery:2014nha,Edery:2015wha,Edery:2018jyp,Oda:2020wdd,Oda:2020cmi,Kamimura:2021wzf,Oda:2021kvx,Edery:2023hxl,Martini:2024tie,Edery:2021bbe},
that the tensor equation of motion of these theories fails
to remain invariant under these transformations (though in~\cite{Edery:2014nha} and~\cite{Martini:2024tie} it was noticed that the trace
of the classical energy-momentum tensor does transform).
This puts into question the sense in which they could
be considered symmetries. Understanding and explaining this discrepancy 
between off-shell and on-shell behaviour was the motivation behind
considering a series of simple examples in 
Secs.~\ref{sec: First example}--\ref{sec: Fourth example}.

The examples we considered all share the feature observed in scalar-tensor theory: 
they possess a global symmetry that, when promoted to a restricted local symmetry, naively 
leaves the action invariant, while transforming the equations of motion.
However, a closer inspection reveals that,
despite the superficial appearance, 
these transformations are typically {\it nonlocal}. 
This is a consequence of the fact that restriction equations often depend on 
dynamical variables of the theory. Consequently,
their solutions exhibit functional dependence on dynamical variables,
that should be accounted for carefully when applied off-shell.

Nonlocal transformations have to be considered with due care when being applied at
the level of the action. Were we to apply the streamlined action principle 
algorithm, valid for local transformations, we would run the risk of getting the equations of motion wrong.
We have indeed shown that nonlocal transformations can generate would-be boundary 
terms that {\it do} contribute to the equations of motion. We have also 
shown that the variations of trajectories of dynamical variables
cannot in general be taken to vanish at the boundaries of the action integral,
after a nonlocal transformation. These boundary variations also {\it do}
contribute to the equations of motion.
Only when these two contributions are both accounted for do we obtain the
correct equations of motion after the nonlocal transformation.

In Sec.~\ref{sec: Scalar-tensor theory} we applied the lessons learned 
from studying simpler examples to the scalar-tensor theory given by
the action in~(\ref{STaction}). The Klein-Gordon equation that the 
conformal factor of the transformation satisfies forces it to be a 
nonlocal function of the metric. Thus, the boundary terms, generated by applying
the transformation to the action, cannot be discarded, nor can the boundary
variations of the fields be neglected. This explains why 
restricted Weyl transformations fail to preserve the equations of motion of the scalar-tensor 
theory: they fail to do so even when applied to the action. We have 
demonstrated this explicitly for the cubic perturbations around Minkowski 
space in Sec.~\ref{sec: Scalar-tensor theory}. Only when boundary variations
are accounted for does the transformed action yield the same equations
of motion as having applied the transformation to the equations of motion 
directly. Thus, we conclude, contrary to the claims in the literature, that 
``restricted Weyl symmetry transformations'' are neither local transformations,
nor are they in fact symmetries.

Restricted subsets of Weyl transformations, such as the 
harmonic Weyl group we considered in this work, and other subgroups~\cite{Edery:2014nha,Kuhnel:1995,ORaifeartaigh:1996hvx,Iorio:1996ad,Martini:2024tie},
are indeed interesting in their own right for their group-like properties.
They are also interesting from the point 
of view of geometry, and the study of transformation properties of curvature 
tensors and other scalar invariants. 
However, even though the scalar densities that make up the action of the
theory might be invariant under such transformations, it is in 
no way guaranteed that the dynamics of the theory is also invariant.
We have shown that one should keep in mind the nonlocal nature of
boundary terms and boundary variations, which can have profoundly 
impact the equations of motion.

If one does not consider the metric as a dynamical field, but rather as 
a fixed external spacetime on which a test scalar field evolves,
as was assumed in parts of~\cite{Martini:2024tie} then only the scalar 
equation of motion remains. That equation by itself is invariant under 
restricted Weyl transformations,\footnote{We are grateful to Omar Zanusso 
for pointing this out.}
that should now be interpreted as mapping test scalar
field solutions propagating on different spacetimes.
When considering theories with a dynamical metric, as we do in this work,
it is still unclear whether or not one can find other realizations 
of ``restricted symmetries'' that would actually be symmetries of the 
equations of motion.
Any such attempt should address the subtleties we pointed 
out in this work, in order to avoid the kinds of inconsistencies encountered 
when promoting global scale invariance to ``restricted Weyl symmetry'',
which does not leave equations of motion invariant.
It might be more
reliable, and perhaps more practical, to consider restricted
transformations directly at the level of equations of motion. 
For example, we note in passing that in the pure~$R^2$ theory an on-shell local 
rescaling~$g_{\mu\nu} \!\to\! R^2 g_{\mu\nu}$ indeed leaves the equation of motion
invariant, provided that~$R\!\neq\!0$. In  this theory, ${\dalembertian} R \!=\! 0$
is true both before and after the transformation. However, understanding
the significance of this observation is beyond the scope of this work.


\begin{acknowledgments}


We are grateful to Paolo Rossi for useful comments on 
the draft of our paper. DG was supported by the European Union and the Czech Ministry of Education, 
Youth and Sports 
(Project: MSCA Fellowship CZ FZU I --- 
CZ.02.01.01/00/22\textunderscore010/0002906).
RN acknowledges GAČR grant EXPRO 20-25775X for financial support.
TZ acknowledges support from the project No. 2021/43/P/ST2/02141 co-funded by 
the Polish National Science Centre and the European Union Framework Programme 
for Research and Innovation Horizon 2020 under the Marie Sk\l{}odowska-Curie grant 
agreement No. 945339.
\end{acknowledgments}


\end{document}